\documentclass{article}
\usepackage{hiph-preprint}
\usepackage{graphicx}
\usepackage{amssymb}
\usepackage{floatflt}
\usepackage{setspace}
\usepackage{subfigure}
\volnumber{22} \issuenumber{1} \edyear{2005}                             
\frompage{000} \topage{000}                                              
\recrevdate{31 October 2005}                                             
\def\rightmark{Nuclear $k_{T}$}
\def\leftmark{T. Henry}

\title{Nuclear $k_{T}$ in d+Au Collisions from Multiparticle Jet Reconstruction at STAR} 
\authors{ 
{Thomas Henry$^{\dagger}$ for the STAR 
Collaboration\footnote{For the full author list and acknowledgments see                               
Appendix {}``Collaborations" in this volume.} %
\index{Henry, T.} 
}\\[2.812mm]
{\normalsize
\hspace*{-8pt}$^{\dagger}$ Texas A\&M University,\\ 
College Station Texas, USA\\[0.2ex] 
}}
 
\abstract{This paper presents the most recent nuclear $k_T$ measurements from
 STAR derived from multiparticle jet reconstruction of d+Au and p+p collisions at 
$\sqrt{s}=200$ GeV.  
Since jets reconstructed from 
multiple particles are relatively free of fragmentation biases, nuclear $k_{T}$
 can be measured with greater certainty in this way than with traditional 
di-hadron correlations. Multi-particle jet reconstruction can also be used 
for a direct measurement of the fragmentation function.}

 
\makeindex
\begin{document}
 
\thispagestyle{empty}
\maketitle

\section{Introduction}\label{intro}

Many analyses of Au+Au collisions at RHIC probe the dense medium using
jets by studying medium 
modifications of the jet yields and jet shapes \cite{Medium Mod}.
The results from d+Au and p+p collisions provide the 
baseline for medium modification studies in Au+Au collisions.
For these simpler systems, jets can be examined
in great detail via full jet reconstruction.

This study presents the most recent d+Au nuclear $k_{T}$ measurements from 
full jet reconstruction at RHIC based upon year 2003 STAR data.  The transverse
and longitudinal jet shapes characterized by the $j_T$, and $z$ distributions 
from p+p collisions are also presented.  
$k_T$ measures the transverse momentum of a
parton within a nucleon or nucleus, and its growth from p+p to nuclear
collisions is believed to contribute to the Cronin effect \cite{cronin}.

\section{Experiment}

The STAR detector is well suited for investigating jet
production at RHIC, due to the complete $\phi$ coverage and large $\eta$
coverage of the TPC and EMC.
The STAR Barrel EMC data used in this study
provide neutral energy measurements including $\pi^{0}$ decay photons. 
During the 2003 RHIC run, only half of the full barrel EMC was installed. This 
first half of the detector consists of 2400 towers of $0.05 \otimes 0.05$ in 
$\eta \otimes \phi$ each, leading to full azimuthal coverage in $0< \eta <1$.
The barrel EMC was read out in minimum bias events. It was also used to 
trigger on {}``high tower'' events, where one of the towers was
above a nominal energy threshold of $E_{T}>2.5$ GeV.
This {}``high tower'' triggered event sample contains a much larger
fraction of jets than the minimum bias event sample.

\begin{figure}[h]
\begin{spacing}{0.8}
\begin{flushleft}\hfill{\includegraphics*[%
  width=0.50\columnwidth]{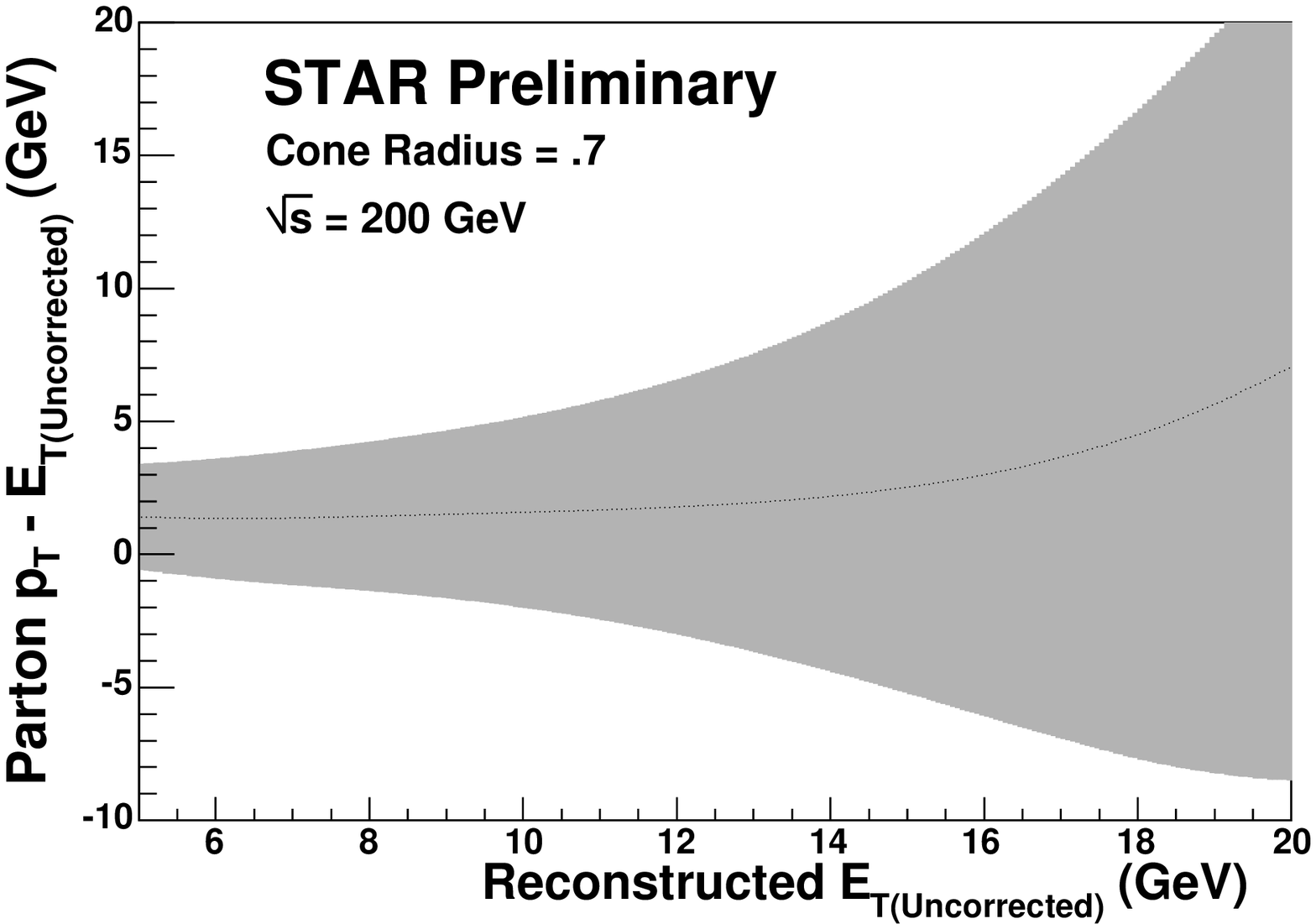}}{\includegraphics*[%
  width=0.50\columnwidth]{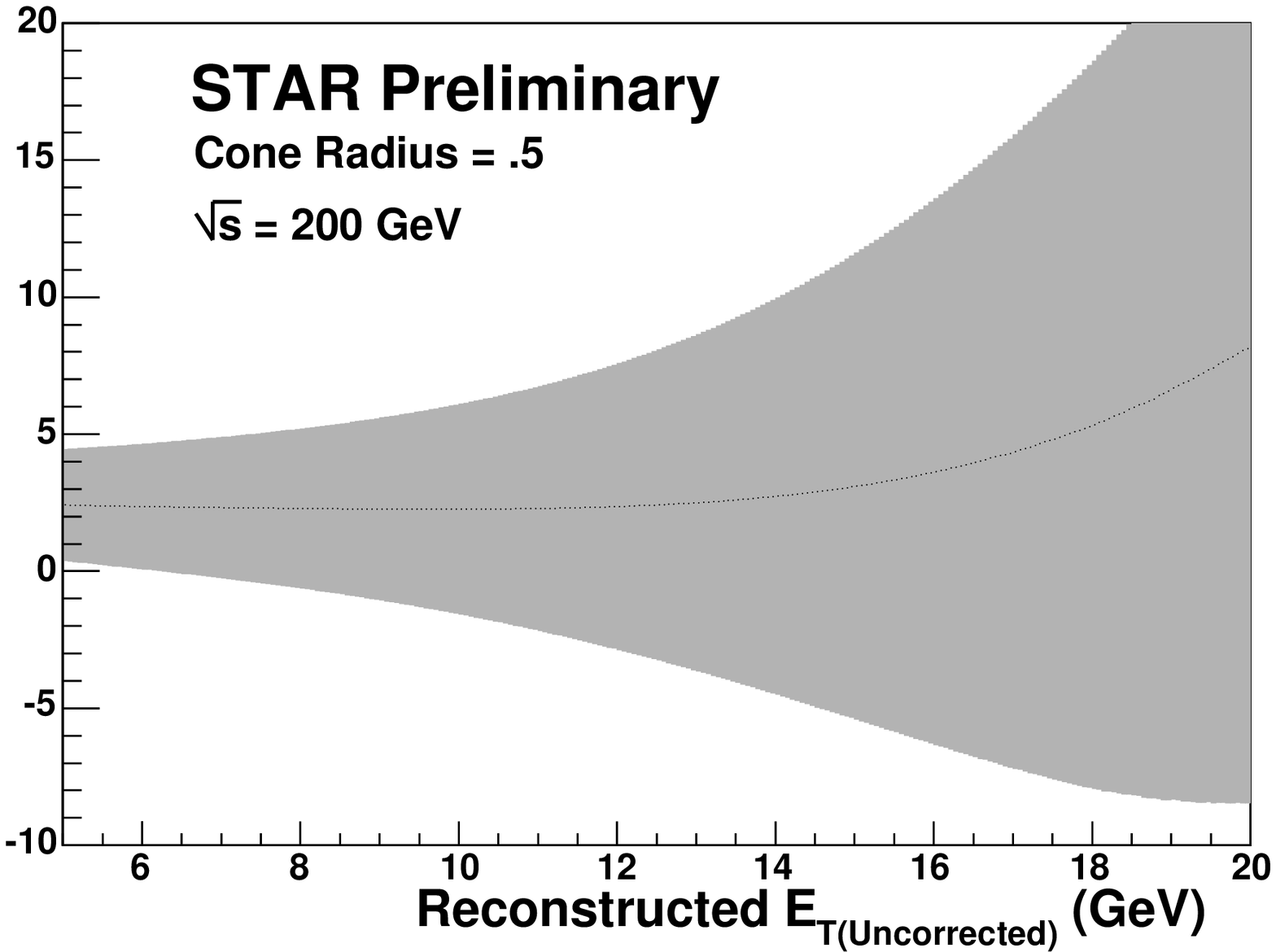}}\end{flushleft}
\vspace*{-0.25in}
\caption{Average difference between the simulated parton $E_{T}$ 
and the total reconstructed jet $E_{T}$.  The gray region outlines the 1-standard 
deviation spread of this average difference.} 
\end{spacing}
\end{figure}

This analysis uses two different algorithms that
reconstruct a jet by capturing the spray of fragmenting particles
in a geometric cone.  One centers the cone on the most energetic hadrons
in the event, while the other optimizes the cone direction to maximize
the included energy. Only the hadron 
centered cone algorithm is robust enough at high multiplicity to be used for
reconstructing jets in d+Au collisions. 

Measurement of jet energy requires corrections for charged particle energy
deposition in the EMC, for the finite efficiencies of the TPC and EMC, 
and for the unmeasured energy carried by long-lived neutral particles 
($n$, $K_L$, ...).
It is also possible for the jet reconstruction algorithm to miss soft 
particles. 

A sample of {\sc pythia}\footnote{PYTHIA version 6.203 and 6.205.} events processed 
by the STAR detector simulator was 
analyzed in the same fashion as the data.  Since the originating parameters 
for each parton were recorded in the monte-carlo {\sc pythia} sample, the energy scale
response of the STAR detector could be quantified.  Figure 1 shows the total 
energy scale correction for jet radii of 0.7 and 0.5 as a function of 
$E_{T(Uncorrected)}$.  From here on, $E_{T}$ will exclusively refer to 
reconstructed transverse jet energy, corrected for detector energy response using 
the curves in Figure 1.

\section{Inclusive Jet Studies}
\begin{figure}[h]
\begin{spacing}{0.8}
\begin{flushleft}\hfill{\includegraphics*[%
  width=0.50\columnwidth]{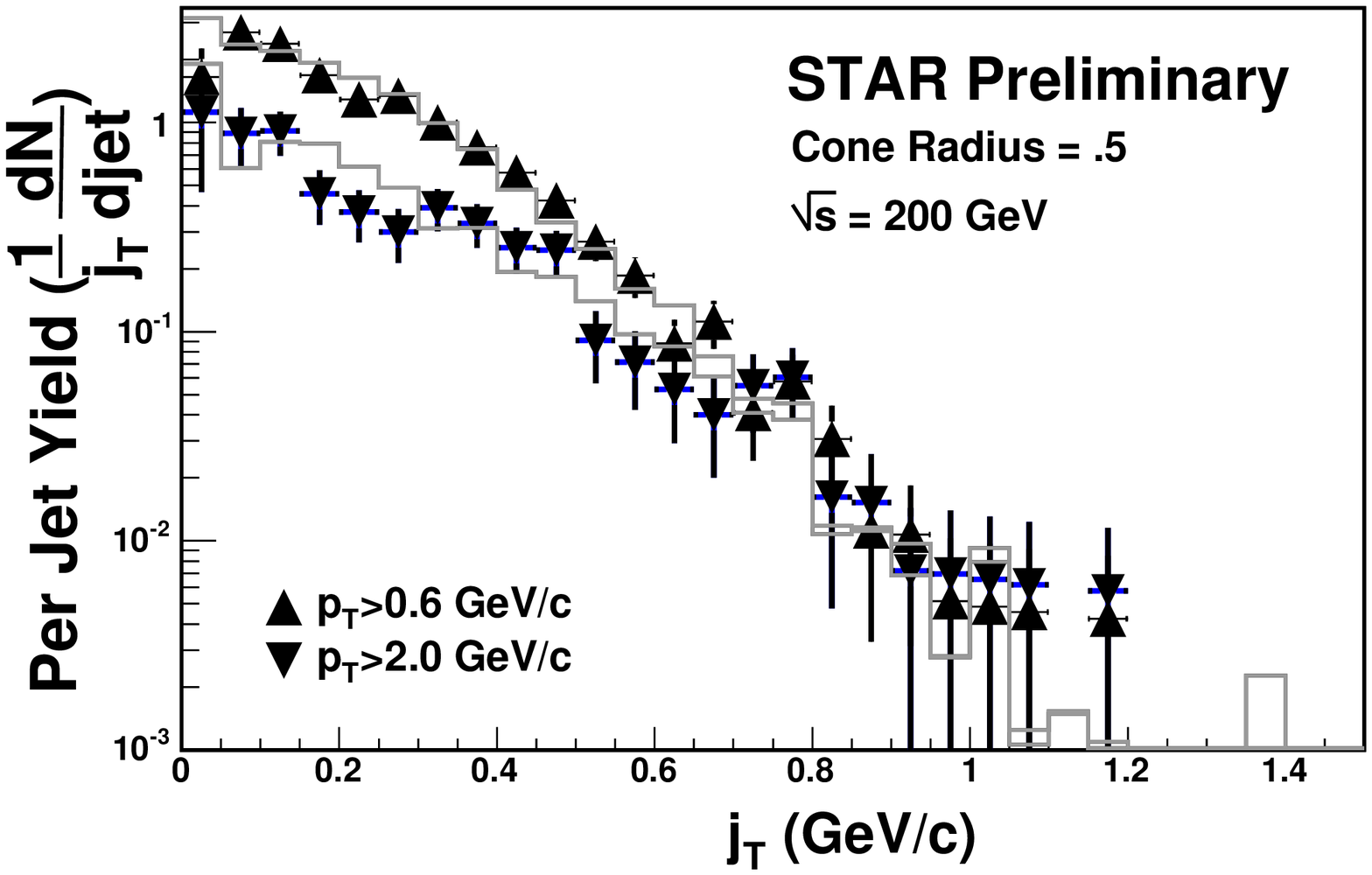}}{\includegraphics*[%
  width=0.50\columnwidth]{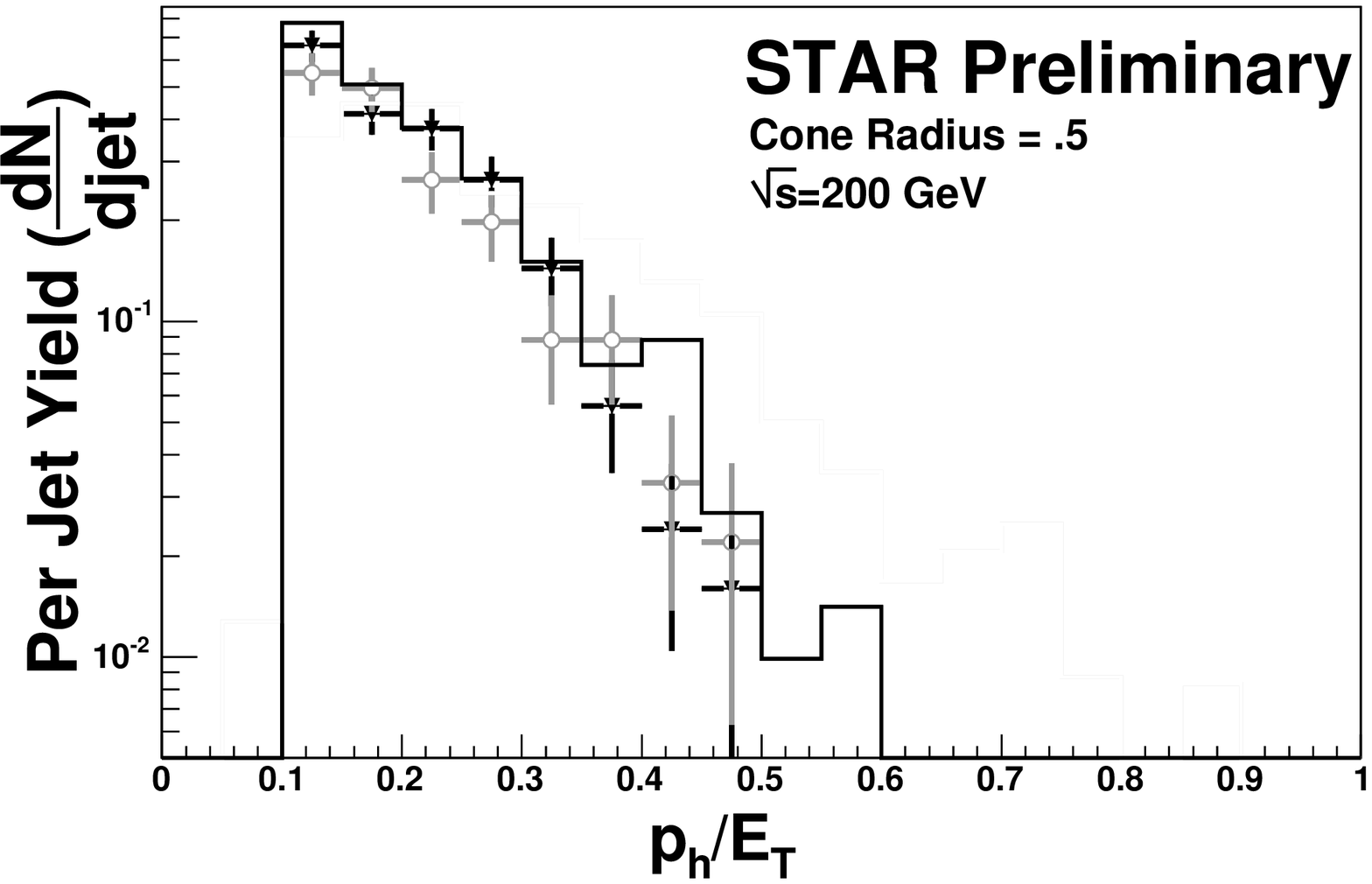}}\end{flushleft}
\vspace*{-0.25in}
\caption{The left panel shows jet $j_{T}$, the transverse component of the 
hadron momentum ($p_{h}$) relative to the jet axis, for charged hadrons.
The upper (lower) data and grey {\sc pythia} monte-carlo histograms refer to 
hadron $p_{h}>0.6 GeV/c$ ($p_{h}>2.0 GeV/c$).  For $p_{h}>2.0$ 
GeV/c, RMS $j_{T} = 486\pm36$ MeV.  The right panel shows $p_{h}/E_{T}$ for 
charged hadrons.  The black data points and monte-carlo histogram are for 
$6.0<E_{T}<7.0 GeV$.  The grey data points are for $7.0<E_{T}<9.0 GeV$.  
The upper range {\sc pythia} monte-carlo curve is statistically consistent with the
lower, and is not shown.} 
\end{spacing}
\end{figure}

Jets in this study are required to pass the following cuts: number charged 
hadrons $>1$, Cone Radius=0.5, $\frac{E_{neutral}}{E_{charged}}<.65$, 
and jet thrust axis $>40(70)cm$ from the inside(outside) of the BEMC 
(roughly $.2<\eta<.65$ depending on z-Vertex).  
When analyzing jet shape, the perpendicular component of the hadron momentum ($j_T$),
and the fragmentation fraction $z$, are generally used.
Figure 2 shows jet $j_{T}$ and $p_{h}/E_T$ distributions using these cuts.
Jets with high hadron count are biased toward higher 
detector energy response.
The $p_{h}/E_{T}$
distribution, since it is a per hadron distribution, is biased toward jets with high
hadron count.    
This means $p_{h}/E_{T}$ is not
directly comparable to the theoretical fragmentation function Dh(z), due to the 
bias introduced by the high hadron count jets.  However, the substantial agreement
between {\sc pythia} and the data demonstrates that the phenomenological parameterizations 
embodied by {\sc pythia} continue to be valid at these jet energies.

The di-jet opening angle is another basic observable obtained from jet 
events. For di-jets, $\Delta\phi=\pi$ in leading order QCD.
Gluon radiation broadens $\Delta\phi$, which
is measured by the per-parton intrinsic 
$k_{T}\equiv E_{T}\sin{\Delta\phi}$.
When reconstructing di-jets, one {}``trigger'' jet
with observed $E_T > 7$ GeV is reconstructed from both neutral and charged 
hadrons using the high tower which triggered the event. 
The other {}``away'' jet direction is reconstructed using charged particles 
only, allowing the {}``away'' jet to range $-0.5<\eta<0.5$.  The {}``away'' 
jet is not used to estimate jet energy scale.

Figure 3 shows the di-jet $k_{T}$ distribution comparison for p+p and d+Au collisions.  
The p+p collisions have been supplemented with d+Au minbias background such that the
multiplicity distributions of the supplemented p+p events matches that of the
d+Au {}``high tower'' sample.
Taking $\sigma_{k_{T}(obs)}^{2}=\sigma_{k_{T}(pp)}^{2}+\sigma_{k_{T}(nucl)}^{2}$,
the $\sqrt{\sigma_{k_{T}(nucl)}^{2}}=1.21\pm0.35\pm^{0.65}_{0.57}$
GeV/c in d+Au collisions. The major systematic uncertainties 
are the jet energy scale in p+p and d+Au, the 
detector resolution, fit uncertainties, and d+Au background multiplicity.  
The fit uncertainties are largest.  The systematic error on the d+Au 
$\sigma_{k_{T}(nucl)}$ due to the detector $k_{T}$ resolution is small.  In
contrast, it is potentially large for the absolute measurement 
of $\sigma_{k_{T}(pp)}$, and this effect is currently under study.

\begin{figure}[h]
\begin{spacing}{0.8}
\begin{center}{\includegraphics*[%
  width=0.50\columnwidth,
  keepaspectratio]{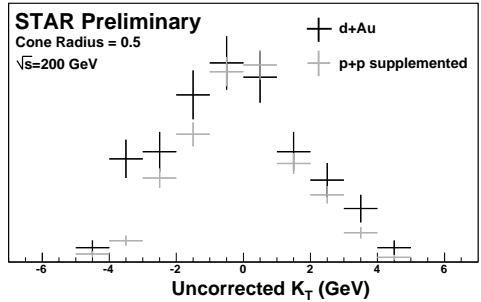}
}\end{center}
\vspace*{-0.25in}
\caption{$k_{T}$ distributions at $\sqrt{s}=200$ GeV.  d+Au $k_{T}$ is shown in black, 
and p+p supplemented with d+Au background is shown in grey. Here $k_{T}$ has been 
corrected for energy but not thrust axis resolution.} 
\end{spacing}
\end{figure}

\vspace*{-0.25in}
\section{Conclusion}
Jets have been reconstructed in both p+p and
d+Au collisions. The p+p jets
have been used to measure $j_{T}$, $z$ and $k_{T}$.
The inclusive jet $j_{T}$ and $z$ 
results compare well with the {\sc pythia} monte-carlo simulation studies.
The d+Au $k_{T_{nucl}}$ value, though having large uncertainty, appears to 
be smaller than previous experiments \cite{E609 Collaboration} at much lower 
energy.
The $\langle j_T \rangle$ and $\sqrt{\langle k_T^2 \rangle}$ results from
full jet reconstruction agree well with similar measurements using
di-hadron correlations \cite{subhasis}.

\vfill\eject
\end{document}